# Conduction Mechanisms in Epitaxial NiO/Graphene Gas Sensors

Somayeh Saadat Niavol * [1,2], Melanie Budde [2], Alexandra Papadogianni [2], Martin Heilmann [2], Hossain Milani Moghaddam [1], Celso M. Aldao [3], Giovanni Ligorio [4], Emil J. W. List-Kratochvil [4,5], Joao Marcelo J. Lopes [2], Nicolae Barsan [6], Oliver Bierwagen [2] and Federico Schipani * [3,6]

[1] Department of Solid-State Physics, University of Mazandaran, 4741695447, Babolsar, Iran
[2] Paul-Drude-Institut für Festkörperelektronik, Leibniz-Institut im Forschungsverbund Berlin e.V., Hausvogteiplatz 5-7, 10117 Berlin, Germany
[3] Institute of Material Science and Technology (INTEMA), Av. Colon 10850, 7600 Mar del Plata, Argentina
[4] Humboldt-Universität zu Berlin, Institut für Physik, Institut für Chemie & IRIS Adlershof Brook-Taylor-Straße 6, 12489 Berlin, Germany
[5] Helmholtz-Zentrum für Materialien und Energie GmbH, Brook-Taylor-Straße 6, 12489 Berlin, Germany
[6] Institute of Physical Chemistry, Eberhard Karls University Tübingen, Auf der Morgenstelle 15, 72076 Tübingen, Germany

## Abstract

Integrated, highly sensitive and reversible sensor devices for toxic and hazardous gases in environmental pollution monitoring can be realized with graphene-based materials. Here we show that, single layer graphene grown on SiC can be utilized to implement sensor devices being extremely sensitive towards $NO_2$ showing an *n-type* response. A second type of sensor with an added NiO layer on top of the single layer graphene changed its response to *p-type* but did not reduce its sensitivity. We show that the conduction switch from *n-type* to *p*-type was not a consequence of an alteration of the graphene layer but is found to be an effect of the NiO layer. We find that the NiO leads to lowering of the Fermi level to a point that a crossing of the Dirac Point in the graphene switched the conduction type. These sensors were tested in the 100 ppb $NO_2$ regime, showing good response and a detection limit extrapolated to be below 1 ppb. This new NiO/graphene/SiC configuration can be an attractive *p*-type sub-ppb sensor platform for $NO_2$ and related gases.

# 1. Introduction

Gas sensors, mostly based on semiconducting metal oxides (SMOXs), are indispensable for safety and health critical applications, e.g. explosive and toxic gas alarms, controls for intake into car cabins, monitoring of industrial processes, as well as future applications such as disease diagnosis via measurement of biomarkers in exhaled breath [1-4].

Since the discovery and isolation of a single molecule layer of carbon in 2004 [5], graphene has rapidly emerged as a highly important material for applications in nanoelectronics, flexible electronics, sensors, and solar cells due to its unique two-dimensional carbon nanostructure that exists in a honeycomb lattice [6]. It is a very attractive material because its entire area can be exposed to gases. Every atom in graphene is a surface atom that can interact with target gases leading to an ultra-sensitive response that can ultimately lead to the detection of very low quantities of gas [7]. Due to the low density of states near the Dirac point, a small change in the number of charge carriers results in a large change in the electronic state which then can easily be measured as a change in resistance of the sensor [8]. For this reason, graphene and its relatives, graphene oxide (GO) and reduced graphene oxide (rGO), alone or in combination with added SMOXs [9-10], are being extensively studied due to their gas sensing properties that are being directed towards environmental monitoring of nitrogen oxides ($NO_x$) [11]. These gases are air pollutant products of fuel burning products that are very harmful to people when inhaled for a prolonged period of time. In particular here we will concentrate on $NO_2$ being the relevant gas for the so-called summer smog.

Graphene can be prepared by various techniques such as chemical vapor deposition (CVD) [12] mechanical exfoliation of graphite [5], chemical exfoliation [13], and *epitaxial* growth [14]. Amongst all these methods, epitaxial growth of graphene on SiC substrate has been proven to be a very practical technique for mass-production of high quality, uniform and wafer-size area graphene particularly for graphene-based nanoelectronic devices and sensors with SiC-based high temperature integrated circuits [6]. The growth proceeds by temperature-induced sublimation of Si atoms from the SiC surface and subsequent formation of a carbon-rich surface that reconstructs to produce graphene directly on an insulating SiC wafer without the need of transferring the graphene onto an insulating substrate [15]. Such graphene layers have an *n*-type electronic behavior due to charge transfer from a carbon-rich buffer layer at the interface between graphene and SiC [16].

Molecular beam epitaxy (MBE) is a well-established and very precise technique for the fabrication of ultra-pure and well-ordered SMOXs thin films with atomic-scale control over structure and composition for sensing applications [17]. There are some reports on modification of graphene and this material serves as a template for MBE growth of metal oxides such as MgO [18, 19], EuO [20], and MoO3 [21]. However, most of the studies have used CVD graphene or highly oriented pyrolytic graphite (HOPG) with only a limited number reporting on the MBE growth of metal oxides on epitaxial graphene/SiC substrates [22]. Furthermore, due to the relatively recent discovery of graphene-based materials as gas sensors, not much fundamental knowledge is available about the sensing mechanisms in such material systems. In this work, we used a new hybrid structure of MBE grown *p*-type NiO film on epitaxial graphene as a sensor. We then propose a model system to explain and explore its behavior and test its sensing capabilities towards $NO_2$ which can potentially show good responses in a sub-ppb concentration range.

## 2. Experimental

### 2.1 Epitaxial graphene growth on SiC(0001)

Epitaxial graphene (EG) was grown on the Si-terminated face of semi-insulating 4H-SiC (0001) substrates (10 × 10 mm$^2$) by employing the surface graphitization method in an inductively heated furnace [23, 24]. First, the substrates were chemically cleaned in an ultrasonic bath by using n-butyl-acetate, acetone, and methanol. The subsequent H-etching was performed at 1400 °C for 15 min in a forming gas atmosphere (95% Ar and 5% H) of 900 mbar and a flow rate of 500 sccm. The H-etching treatment is performed in the same furnace where graphene is subsequently synthesized. Finally, epitaxial graphene was prepared by silicon sublimation of the substrate at 1600 °C in a 900 mbar Ar atmosphere under a flow rate of 500 sccm for 15 min [24].

### 2.2 NiO deposition

NiO layers were grown by MBE in an ultrahigh vacuum (UHV) system with low base pressure of $10^{-10}$ mbar onto single crystalline graphene layers and co-loaded SiC(0001) substrate without graphene as reference. Molecular oxygen and a nickel effusion cell were used for the growth. The temperature was kept well below the melting point of nickel (1455 °C), at 1380 °C in order to protect the nickel effusion cell. The growth temperature is defined as the substrate heater temperature, measured by a thermocouple between substrate and heating filament [25]. An elemental Ni flux was introduced to both substrates maintained at 100 °C for 35 min. The growth of a nominally 10 nm thick NiO layer was achieved under oxygen rich conditions by introducing molecular oxygen (0.5 sccm) after one minute into the UHV chamber. The nominal NiO thickness was calibrated by growth of NiO films on MgO(100) [25]. The NiO film thickness on SiC and graphene has also been confirmed by analyzing the fringe-spacing in x-ray reflectivity (XRR) and x-ray diffraction (XRD) measurements around the symmetric out-of-plane NiO-layer reflex (not shown here). Both methods, whose application to NiO on MgO(100) has already been demonstrated by us,[25] are described in detail in Ref. [26]. As a result, the smooth NiO film on SiC has a thickness of 9.8(+/-0.5) nm whereas the NiO film on graphene (excluding tall islands) is only 8.4(+/-1.1) nm thick. The discrepancy between both films is due to the tall NiO islands on graphene (will be confirmed by atomic force microscopy in the next section) that do not contibute to the fringes in XRD.

### *2.3 Atomic force microscopy*

We used atomic force microscopy (AFM) to determine the surface morphology of graphene, NiO/SiC and NiO/graphene/SiC samples. The morphology of the pristine epitaxial graphene used in this study (Fig. 1a) is composed of step edges and terraces with widths of about 5.2 μm, which form due to the step bunching process during epitaxial growth and whose dimensions rely on the miscut of the SiC wafer [24]. The step edge regions are covered by few-layer graphene (usually bi- or trilayer) and the flat terraces mainly by single-layer graphene [27, 28]. Fig. 1b and Fig. 1c exhibits the AFM height and phase images of NiO/graphene/SiC sample. Although not clearly visible in the height image (Fig. 1b), the corresponding the phase-contrast image (Fig. 1c) reveals a grainy surface morphology in the dark (lower lying) areas of the height image. Considering the atomically smooth nature of epitaxial graphene [24, 27-29], this result strongly suggests the complete coverage of the graphene surface by NiO (confirmed later by XPS). Despite this, two phase-contrasts are observed in Fig. 1c. This effect can be attributed to the large height difference between these areas, which is known to contribute to phase-contrast in AFM [30]. From the growth point of view, the height variation observed in NiO might be related to distinct growth behaviors taking place in graphene areas exhibiting different thickness, similar to what has recently been observed for MBE synthesis of hexagonal boron nitride on graphene/SiC(0001) [31]. More investigations to elucidate such aspects are under progress. The height image of NiO/SiC sample (Fig. 1d) reveals that the smooth surface of SiC is fully covered by the NiO layer. The NiO grains on graphene (rms roughness =2.33 nm) are significantly taller than those on SiC (rms roughness =0.46 nm).

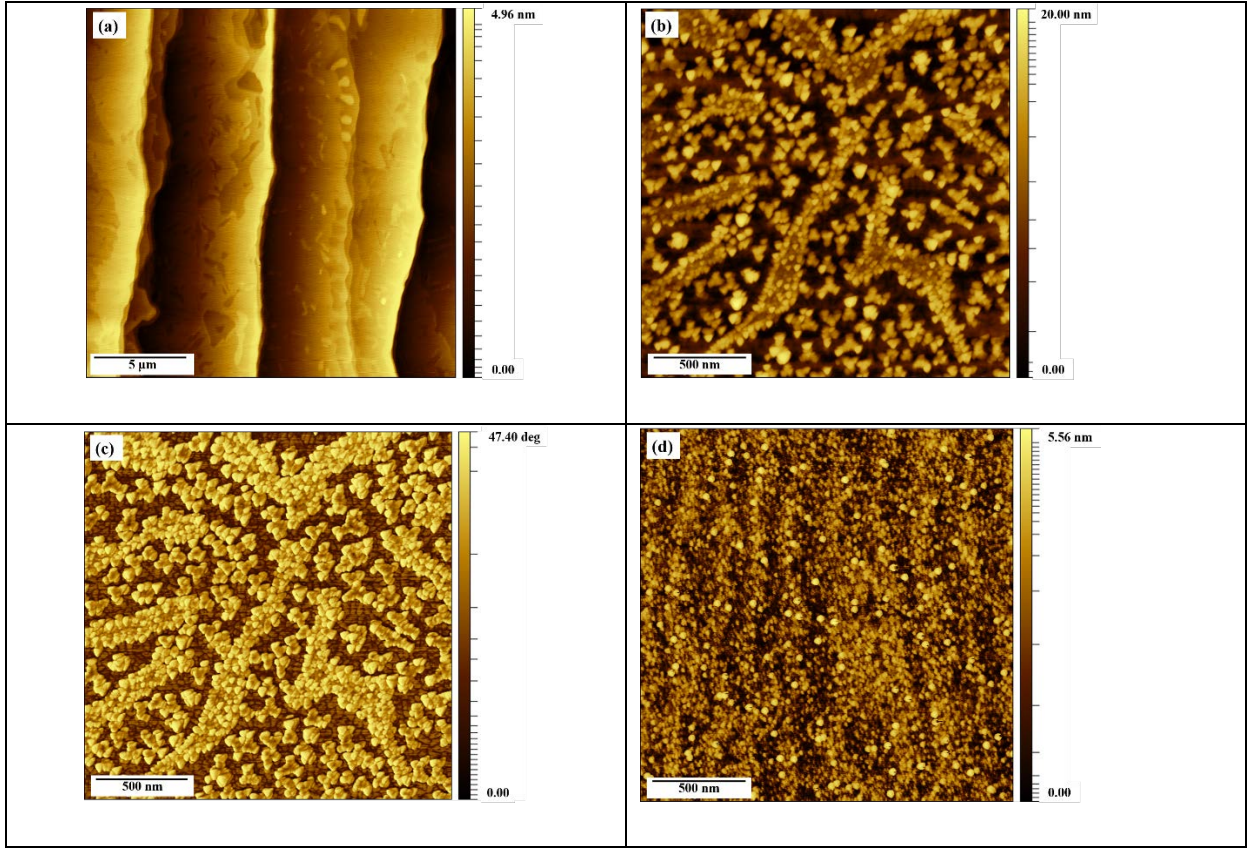

**Figure 1**: (a) AFM height image of pristine graphene prepared on H-etched 4H-SiC (0001) using the surface graphitization method over an area of $20 \times 20$ µm$^2$ , (b- height; c- phase contrast ) images of NiO/graphene/SiC over an area of $2 \times 2$ µm$^2$, and (d) AFM height image of NiO /SiC over an area of 2×2 µm$^2$.

### *2.4 Raman spectroscopy*

Raman spectroscopy was performed to investigate the structural quality of our epitaxial graphene before and after deposition of NiO layer. All measurements were performed at the surface terraces (approximately 5.2 µm wide, as determined by AFM), without any contribution of the few-layer graphene existing at the step edges. The contribution of the bare SiC substrate has been subtracted from the original Raman signals. Fig. 2a shows the Raman spectrum of the as-grown graphene obtained using a 473 nm laser source. It exhibits the prominent characteristic G (symmetric $E_{2g}$ phonon mode) and 2D (double resonant electron-phonon process) peaks of graphene at ~1591 cm$^{-1}$ and ~2717 cm$^{-1}$, respectively [29, 32]. The narrow full width at half-maximum (FWHM) of 18.1 cm$^{-1}$ and 31.4 cm$^{-1}$ for the G and 2D peaks shows the existence of a high-quality monolayer graphene on top of SiC [33]. Fig. 2b displays the Raman spectrum after coating the graphene with the NiO layer. Apart from a slight reduction of the 2D peak intensity, there are no significant differences between the spectra especially in the 2D and G peak positions. The intensity of the disorder-induced D peak (at approx. 1350 cm$^{-1}$) is very low, or negligible, as compared to the sharp G and 2D peaks for both samples, indicating that there is only a small density of defects/disorder in the graphene structure before and after deposition of NiO layer [32]. These results demonstrate that the MBE growth of NiO did not degrade the quality of the underlying graphene.

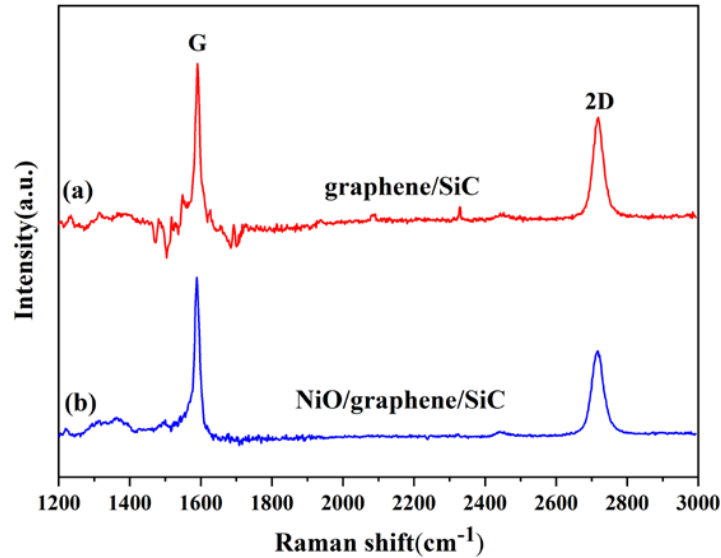

**Figure 2**: Raman spectra of (a) pristine epitaxial graphene and (b) as well as after deposition of NiO on top of graphene.

*2.5 X-ray photoelectron spectroscopy*

X-ray photoelectron spectroscopy (XPS) was performed in a JEOL JPS-9030 photoelectron spectrometer system using the Al Kα (1486 eV) excitation source (a monochromator was employed). The samples were grounded during the photoemission measurements. Fig. 3a and Fig. 3b display the XPS survey spectra of graphene on SiC before and after the growth of NiO. Pristine epitaxial graphene shows a near-surface composition of C, Si and O atoms (Fig. 3a). As shown in Fig. 3b, the XPS survey spectra of NiO layer grown on epitaxial graphene clearly indicate the predominant presence of C, Ni and O. The absence of detectable Si peaks (Si 2s and Si 2p) after the deposition of NiO confirms the full coverage of epitaxial graphene surface by the NiO layer, in agreement with the AFM results. High-resolution Ni 2p and O 1s regions of NiO/graphene sample are shown in Fig. 3c and Fig. 3d. The Ni 2p spectrum exhibits two peaks at binding energies of ~853.7 eV and ~872.8 eV with a separation of 19.1 eV, which agree with the Ni 2p3/2 and Ni 2p1/2 spin orbit levels of NiO [34]. Moreover, a satellite peak of 2p3/2 is observed at ~861.2 eV [35]. The O1s spectrum exhibits peaks at 529.1, 530.5 and 532.1 eV corresponding to O-Ni ($O^{2-}$), O-H and O-C.

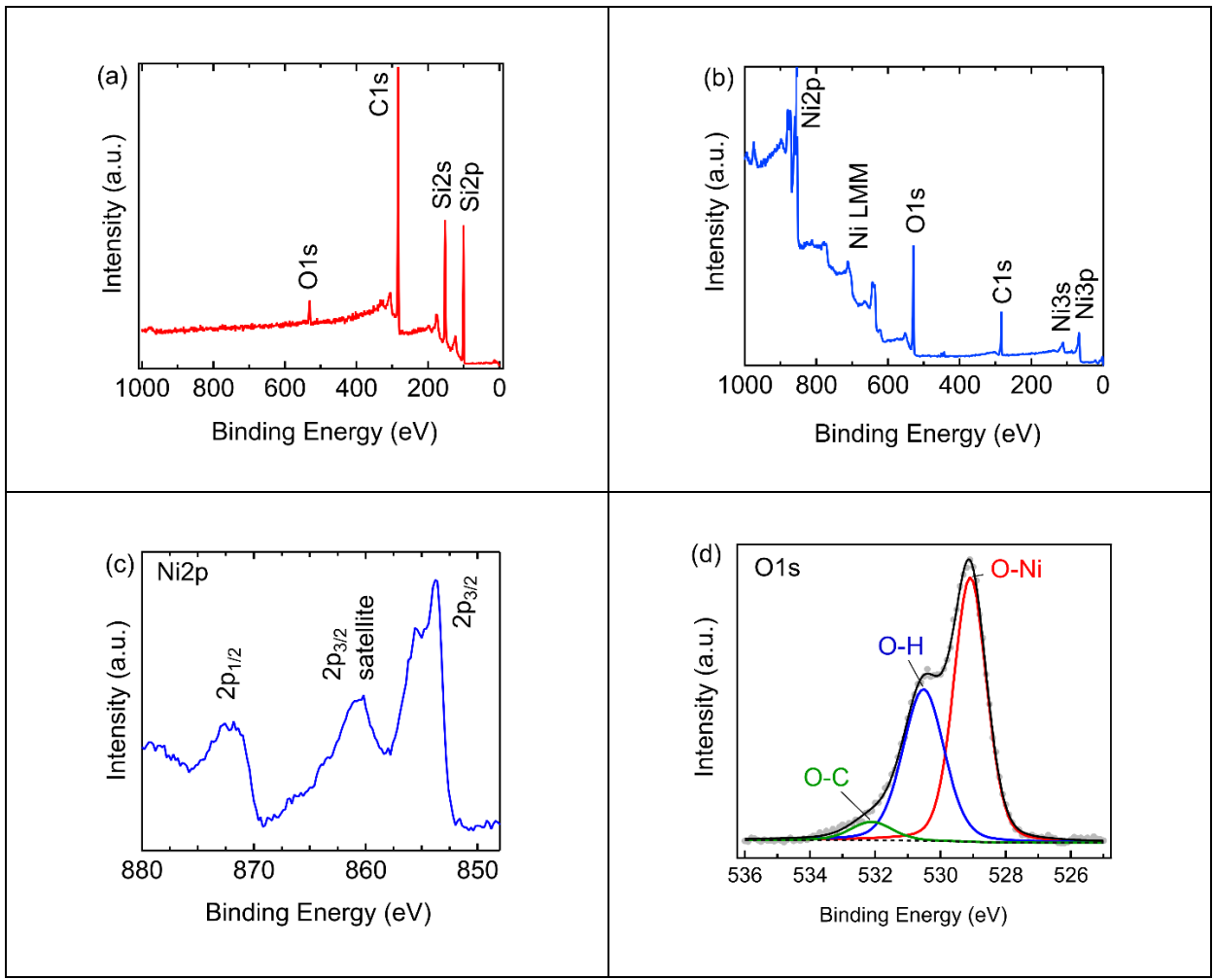

**Figure 3**: XPS spectra of (a) epitaxial *graphene on SiC* and (b) NiO layer grown on epitaxial graphene, (c, d) High resolution XPS spectra of Ni2p and O1s regions of NiO/graphene/SiC sample.

## *2.6 Sensor fabrication*

For the sensor-gas-response measurements, a 4 mm x 7 mm large chip was cleaved out of the samples. In order to provide ohmic contacts to the layer, 0.5 mm diameter disk-shaped contacts of 20 nm Ti/100 nm Au were deposited through a shadow mask in a 2 mm x 2 mm square arrangement close to one of the short edges of the chip by electron-beam evaporation. Subsequently, the chip was fixed with fast-drying silver paint onto a ceramic carrier with integrated heater and the two ohmic contacts closest to the edge were wire-bonded to the electrodes on the carrier to enable *in-operando* two-point resistance measurements.

## *2.7 Transport measurements*

In order to examine the carrier transport within the samples, room temperature (RT) Hall measurements in ambient atmosphere were performed using the four-point van-der-Pauw (vdP) arrangement immediately after growth (as-grown) on the 10 mm × 10 mm large square shaped sample pieces, as well as on the cleaved 4 mm × 7 mm sample after gas-response experiments. The electrical probes were placed directly on the surface of the films at the edges of the samples and for the Hall measurement a current of 0.1 mA and a magnetic field intensity of 0.5 T were applied. Tabs. I and II summarize the results of the transport measurements for the pure graphene and NiO/graphene samples respectively.

|  | As-grown | After gas exposure |
|---|---|---|
| Carrier type | *n*-type | *n*-type |
| $n_{2D}$ ($10^{12}$ cm$^{-2}$) | 3.14 | 2.17 |
| $R_S$ (kΩ/sq) | 2.89 | 4.56 |
| $\mu$ (cm$^2$/Vs) | 688 | 505 |

*Table I: Charge carrier transport properties of the pure graphene sample with $R_S$ being the sheet resistance, $n_{2D}$ the sheet electron concentration, and $\mu$ the Hall mobility.*

|  | As-grown | After gas exposure |
| --- | --- | --- |
| Carrier type | $n$-type | $p$-type |
| $n/p_{2D}$ ($10^{12}$ cm$^{-2}$) | 47.3 | 5.41 |
| $R_S$ (kΩ/sq) | 0.33 | 2.31 |
| $\mu$ (cm$^2$/Vs) | 396 | 499 |

*Table II: Charge carrier transport properties of the NiO/graphene sample with $R_S$ being the sheet resistance, $n_{2D}/p_{2D}$ the sheet electron/hole concentration, and $\mu$ the Hall mobility.*

## 3. Results and Discussion

For the analysis of the sensing properties, band structure, and gas detection limits, three kind of samples were measured, all grown on SiC(0001): NiO, graphene, and NiO/graphene. All were measured at 100 °C.

First, Fig. 4 shows the two-point DC resistance response of the NiO/SiC sample. Nickel oxide can be insulating or have a $p$-type conduction depending on the growth conditions [36], the latter is achieved in this case. The sample was exposed to several target gases, but only showed appreciable response to CO. The $p$-type response indicates that Ni vacancies dominate, behaving like acceptors and the low conductivity is due to a low concentration of these defects. The main purpose of this measurement is to show the behavior of the NiO/SiC sample for later comparison with the other two.

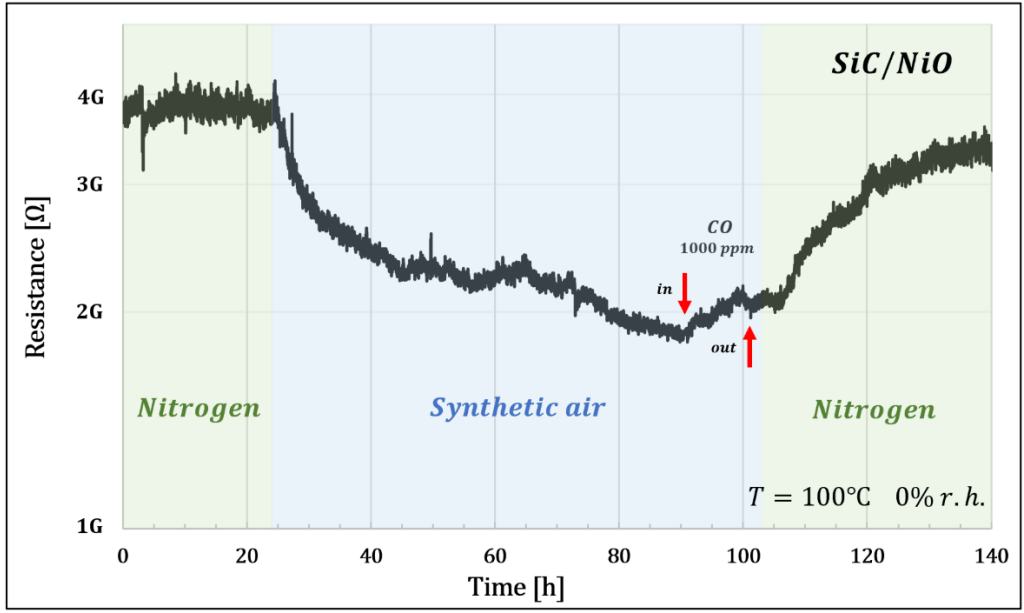

**Figure 4**: Gas response of the NiO/SiC sample. It shows a p-type response both when exposed to an oxidizing gas (O₂) and a reducing gas (CO).

With the intention of drawing a band diagram, we calculated the position of the Fermi level of the sample: since the substrate is insulating, lateral transport dominates the conduction between the two metallic contacts. Therefore, the measured two-point resistance corresponds to the sheet resistance $R_{2p} = 0.97 * R_s$, being $R_s$ the resistance of a square-shaped piece of the film, and 0.97 a geometric factor derived from finite element method simulations [37]. The corresponding sheet resistance of the NiO/SiC sample in synthetic air of ≈ 2 GΩ is in good agreement with that measured by interdigitated contact structures on MBE grown MgO/NiO layers in synthetic air [25]. Using the film thickness of $d$ =10 nm, the corresponding resistivity $\rho$ of the layer can be calculated as $\rho \cong R * d = 2x10^9 \Omega * 1x10^{-8} m = 20\ \Omega m$.

From the conductivity $\sigma = 1/\rho$, expressed in terms of the hole mobility of NiO as

$$\sigma = pe\mu, \tag{1}$$

with $\mu$ between $0.01 - 0.0001\ cm^2/Vs$ [38], the hole concentration can be found to be in the range of $p = 3x10^{17} - 3x10^{19}\ 1/cm^3$. Now we can estimate the position of the Fermi level knowing that

$$p = 2(2\pi m^* k_B T/\hbar^2)^{3/2} e^{-(E_F-E_V)/k_B T}, \tag{2}$$

with an effective mass of $0.9m_0$ [39].

Finally, the Fermi level position can be estimated to lie within a range between the top of the valence band of NiO and 0.15 eV above it. For the sake of simplicity, in this range we will adopt an arbitrary value of 0.05 eV. Regardless of our choice, the Fermi level is close to the middle of the gap of SiC. Band diagrams can be seen in Fig. 7. From the estimated hole concentration, the Debye Length $L_D$ can be found to range between 2,4 – 265 nm and since the width of the NiO layer is 10 nm, the effect of the adsorbed gas is most likely felt through the entire layer. When the sample is exposed to an oxidizing atmosphere, acceptor levels on the NiO surface located below the $E_F$ capture electrons lowering the Fermi level, therefore decreasing the overall sample resistance.

Following NiO/SiC measurements, the graphene/SiC and NiO/graphene/SiC samples were studied at 100 °C and exposed to:
- 100, 250, 500, 1000 and 2500 ppm of $O_2$ in $N_2$ then,
- 5, 10, 20, 50 and 100 ppm of $H_2$ in nitrogen.
- 10, 20, 50 and 100 ppm of CO and finally
- 0.1, 0.2, 0.5, 1, 2, 5, 10 and 20 ppm of $NO_2$ in a background of synthetic air.

Electronically, graphene is a 2D zero-gap material with conically shaped valence and conduction bands [40]. The point where these two bands cross is called Dirac point and for an ideal graphene layer, the Fermi level is located at this intersection. However, epitaxial graphene on SiC(0001) typically shows an *n*-type doping [14] due to charge transfer from the buffer layer below it, which is partially $sp^3$-bonded to the SiC [41]. The reported electron affinity of 4H-SiC is 3.24 eV [42] and it is lower than the work function of free-standing single layer graphene, which is reported to be 4.8 eV [43]. Therefore, there will be a net flow of electrons to the graphene layer that will remain negatively charged. For an *n*-type graphene, the Fermi level lies above the Dirac point inside the conical conduction band (Fig. 7b). The opposite is valid for *p*-type graphene where the Fermi level is inside the valence band, below the Dirac point (Fig. 7c).

The *n*-type electrical behavior of the graphene/SiC sample is corroborated through (by) gas sensing where Fig. 5 shows the response to all tested gases. The first resistance drop is due to a temperature increase from RT to 100 °C. A notable response to $NO_2$ can be observed for concentrations between 2 to 20 ppm where more than half of the final response is achieved in the first minute of exposure. The detection limit to this gas is explored later on the paper.

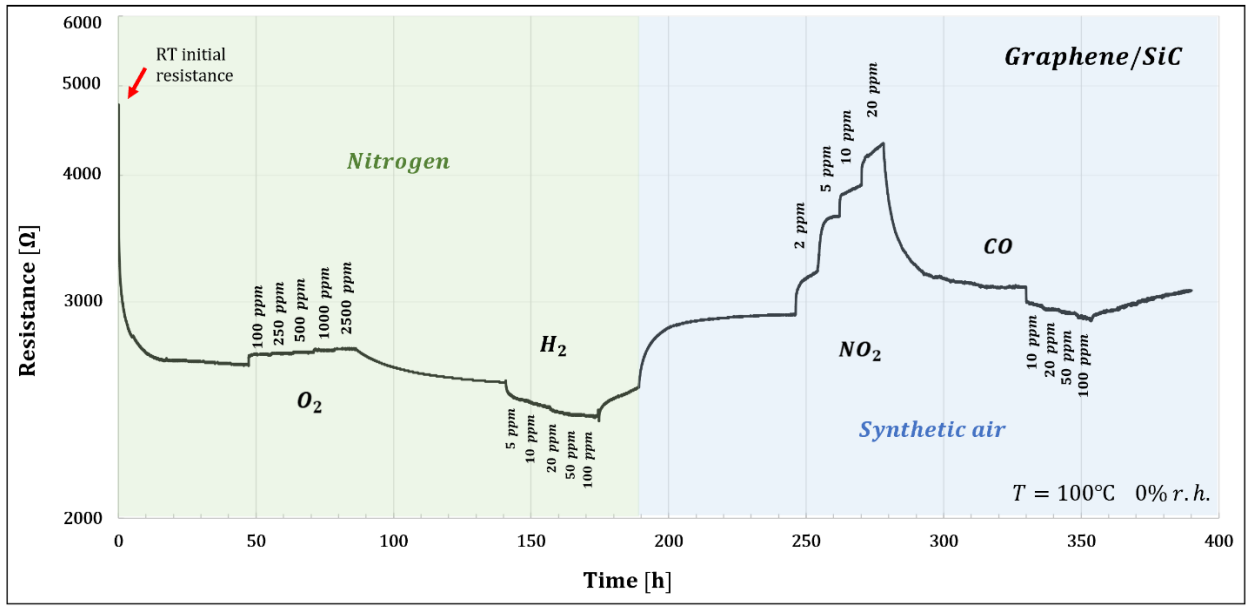

**Figure 5**: Gas response of the graphene/SiC sample.

Finally, a NiO layer on top of a graphene sample was added and measured. As shown in Fig. 1b and Fig. 1c, the NiO layer consists of an approximately 10 nm thick closed layer, covering the entire graphene surface. As seen in Fig. 6, the sensor shows an *n*-type response at first, but after the first exposure to $NO_2$ the sample permanently changes its behavior to *p*-type, decreasing its resistance against oxidizing target gases and the opposite for reducing ones. After a higher temperature treatment in nitrogen (150 °C in this case, not shown here) for at least 24 h, the initial *n*-type response is recovered, and the *n*-to-*p* conduction type switch can be achieved again. The initial *n*-type conductivity of the NiO/graphene/SiC sample and its change into *p*-type after the annealing and gas exposure were also confirmed by Hall effect measurements (see Tab. II).

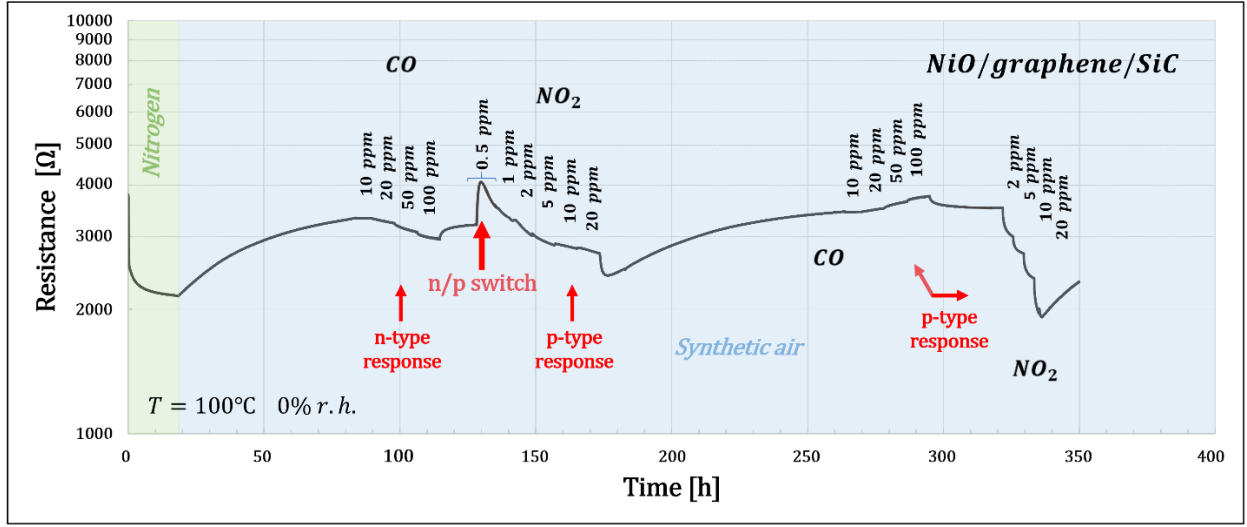

**Figure 6**: gas response of the NiO/graphene/SiC sample at 100 °C and with no humidity present. There is an n/p conduction type switch when exposed to $NO_2$ that is only reversible when the sample is treated at higher temperatures in an $N_2$ atmosphere.

As a first approach, band diagrams can be built following Anderson's rule [44]. Given that the measured resistance is in the order of a few kΩ (in contrast with the few GΩ of the NiO/SiC sample), it is clear that we are measuring the resistance of the graphene layer with no contribution of the NiO to the overall response. Therefore, since the current flows through the path of least resistance, it will flow vertically through NiO and into the graphene, laterally through the graphene and vertically into NiO again to reach the other metallic contact. Taking into account the area of the contact $0.002 cm^2$, the vertical resistance can be estimated as $R_{NiO} = \rho * d/A = 2000 \Omega cm * 1 x 10^{-6} cm / 0.002 \, cm^2 = 1 \, \Omega$.

As seen in Fig. 6, after the first exposure to $NO_2$, the resistance increases for a period of several hours, then decreases. This creates an *n/p* conduction type switch that has been observed and reported by several authors on epitaxially grown graphene on SiC [45, 46]. Since we were not able to obtain a *p*-type behavior from the graphene/SiC sample regardless of $NO_2$ exposure, we conclude the switch is not due to graphene, but a consequence the of NiO layer grown on top, which lowers the Fermi level by a small amount (~ 0.1 - 0.2 eV) enough to cross the Dirac Point of the graphene and become *p*-type. A possible reason for this change of the Fermi level could be due to the appearance of dipoles at the NiO/graphene interface.

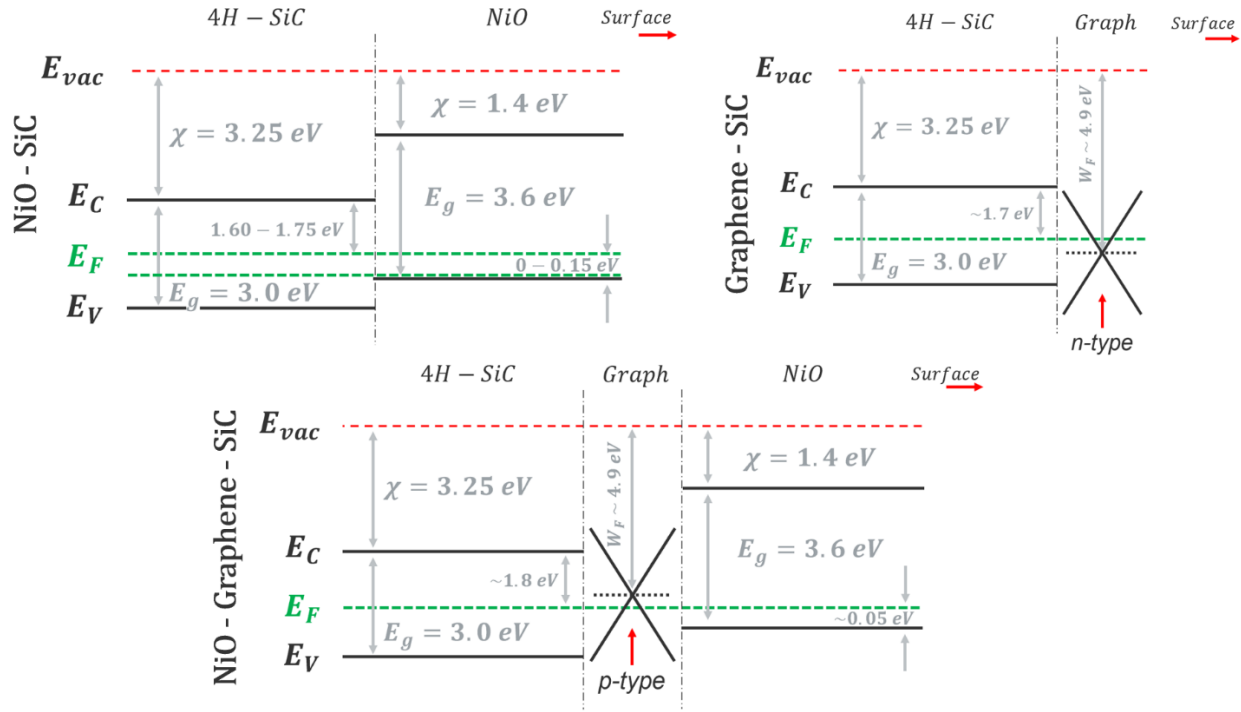

**Figure 7**: band diagram of the samples. a) NiO/SiC interface that shows p-type behavior. b) graphene/SiC interface where the Fermi Level lies above the Dirac Point (DP) of the graphene, therefore always behaving like an n-type material. c) after $NO_2$ exposure, the Fermi level crosses the DP in the NiO/graphene/SiC sample and changes to a p-type response. The used $W_F$ for the graphene is 0.1 eV larger than reported.

The quality of graphene was verified before and after the gas sensing experiments via Raman spectroscopy and no change could be observed (not shown here). Therefore, the measured response is not the result of sample degradation.

Fig. 8 shows the baseline of the NiO/Graphene/SiC sample over time in synthetic air. The orange dots represent its response to 20 ppm of $NO_2$. Response #1 (also shown in Fig. 6) and #5 represent an n/p conduction type switch were resistance first increase to then decrease when exposed to $NO_2$. As stated before, this switch is a consequence of a previous nitrogen exposure. Successive $NO_2$ cycles found good repeatability with a decrease in response of around 20 – 25 % after four consecutive cycles (#6 to #9).

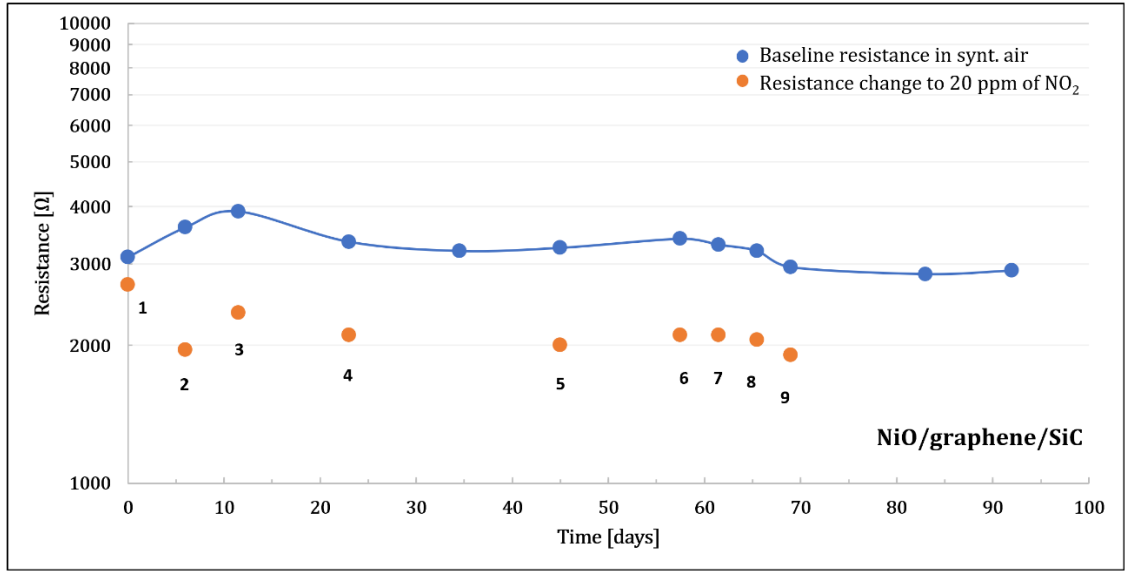

**Figure 8:** baseline resistance over time in synthetic air of the NiO/Graphene/SiC sample. The orange dots represent the response to 20 ppm of NO$_2$ where cycle #1 and #5 are from an n/p conduction type switch due to a previous nitrogen exposure. After four consecutive cycles (#6 to #9) the sensor shows good repeatability only changing its response around 20 - 25 %.

When the sample is exposed to NO$_2$ for extended periods of time (one complete cycle can last up to 56 hs, as seen in Fig. 9a) the recovery time can be up to four days. This is usually not how sensors are exposed to target gases. Therefore, Fig. S1 (in supplementary material) show the response to 2 ppm of NO$_2$ in synthetic air for an exposure time of one hour. We found that 25 % of the total response time is achieved within 1 min of exposure and the baseline recovered in 4 to 5 hours. Consistent responses are found with higher and longer concentration exposure.

Work function measurements were also combined with simultaneously performed two-point DC- resistance measurements. Work function changes were measured with the Kelvin Probe technique, which is a non-contact, non-destructive method that uses a vibrating reference electrode and measures the changes of the contact potential difference (CPD) between the sample and the electrode. Variations in the CPD induced by the changes in the gas atmosphere represent relative work function variations of the sample:

$$-\Delta CPD = \Delta W_F = \Delta\chi + \Delta eV_s + \Delta(E_F\text{-}E_V) \qquad (3)$$

A detailed description of work function and Kelvin Probe working principles can be found in [47].

Since the Kelvin Probe is measuring the NiO surface, an increase in $W_F$ could be explained by the increase of electron affinity $\chi$, that can be attributed to the highly oxidizing NO$_2$ gas [48, 49]. It is also possible that the third term of Eq. 3 became even smaller, but this is a minor contribution to the total change. This change in $\chi$ is purely superficial and does not affect the graphene/NiO interface.

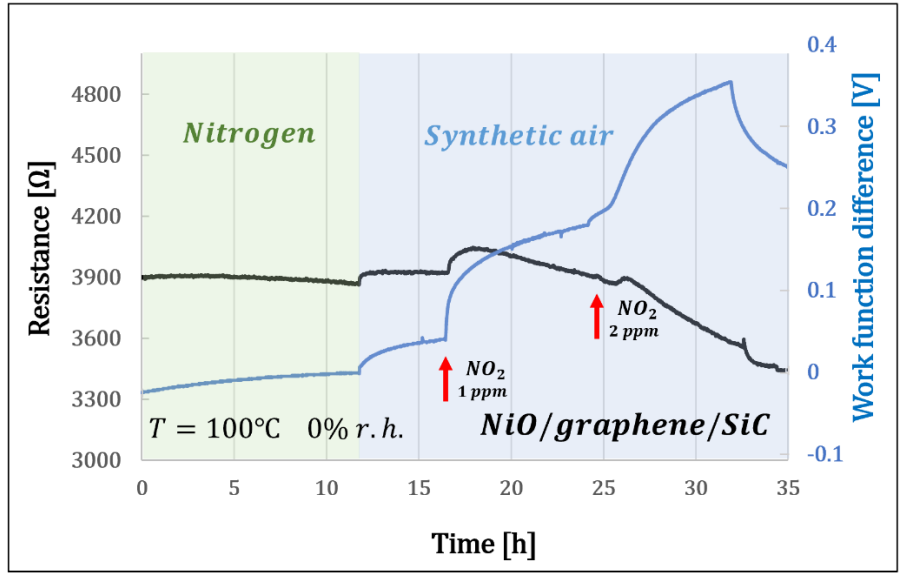

**Figure 9**: Two-point DC resistance and work function measurements.

The NO$_2$ sensing mechanism has been studied in detail by Rosso *et al*. [49] with direct measurement techniques under *operando* conditions on the surface of a SMOX. At low operation temperatures, such as our case, they found that the formation of nitrites at the surface irreversibly increase and accumulate due to the impossibility to form nitrates. However, the impact on the response of this accumulation is too small and not dominant in the sensing process at low temperatures. Contrary to this, hydroxyl groups and water at the surface facilitate NO$_2$ adsorption and are mainly responsible for the response. The slow decrease in work function (most likely a decrease in electron affinity, as mentioned above) and slow electrical response recovery after the NO$_2$ is removed shows that nitrite sticking is most likely present at the surface of NiO.

### 3.1 Detection limit

Graphene is electronically, a very low noise material, making it a good candidate to sense very low quantities of gases. Here, we intend to explore the detection limit of the sensor to NO$_2$.

The main limitations are the reliability of the gas flow to accurately transport very low quantities of a target gas (in the low ppb range), the response of the reference sensor we use, a FIGARO TGS2600, to check that there is effectively a target gas being injected into the measuring chamber, and, most importantly, the signal-to-noise ratio. Therefore, the detection limit was chosen to be whatever target gas concentration produces a signal change at least three times the amplitude of the noise, which is approximately 1/1000 of the signal for both samples or, 100 ppb of $NO_2$ delivered by our gas mixing system, which is a concentration that we are positive to being able to deliver with ± 10 % of accuracy and detect with our reference sensor. Then some extrapolations can be made, if necessary, to show a potential lower detection limit.

Humidity in the system was measured with a Vaisala DRYCAP Hand-Held Dewpoint Meter DM70 that reached its detection limit of 10 ppm in a few hours. All bottled gases used were 6N purity, which means there is a contamination of maximum 1 ppm of water or non-wanted gases. Therefore, humidity in the system is estimated to be between 1 and 10 ppm (0.0001 - 0.001%), which is very low in general, but it may be significant for the lowest target gas concentrations.

Fig. 10a shows the response of the graphene and the NiO/graphene/SiC sample to different $NO_2$ concentrations in a synthetic air background at 100 °C. The percentage of change is shown in Fig. 10b, comparing changes against its baseline resistance. The response of both samples to 0.1 ppm of the gas is very clear, changing a 3.2 % and a 3.7 % for the graphene/SiC and NiO/graphene/SiC respectively. Finally, for the highest concentration, 10 ppm in this case, the trend reverses, and pure graphene shows a higher response: 59 % increase for graphene and 33 % resistance decrease for NiO/graphene/SiC. Given the signal change for the lowest concentration, there is potentially a lower detection limit than experimentally shown. To estimate this, the resistance variation $\Delta R/R$ was plotted against the concentration of $NO_2$, where the potential minimum response is given by three times the average noise divided by the signal. Therefore, a signal change below 0.15 % will be considered noise and not the response to a target gas concentration exposure. According to the fit, both samples show the potential to detect sub-ppb levels of $NO_2$. One has to consider that these measurements were done in very controlled conditions of temperature and humidity but the sub-ppb detection trend that other authors have published [45, 46, 50 - 54] is also corroborated by this work. The main difference is that this sensor has a p-type response to $NO_2$ (due to the NiO on top) and not n-type, as most epitaxially-growth graphene-based sensors reported. Regarding its p-type response, we present the fifth lowest estimated detection limit for all graphene-based sensors reported to our knowledge, and the second lowest for epitaxial graphene-based ones. A table is available as Supplementary Material that sums some of the most important articles published on gas sensing with graphene.

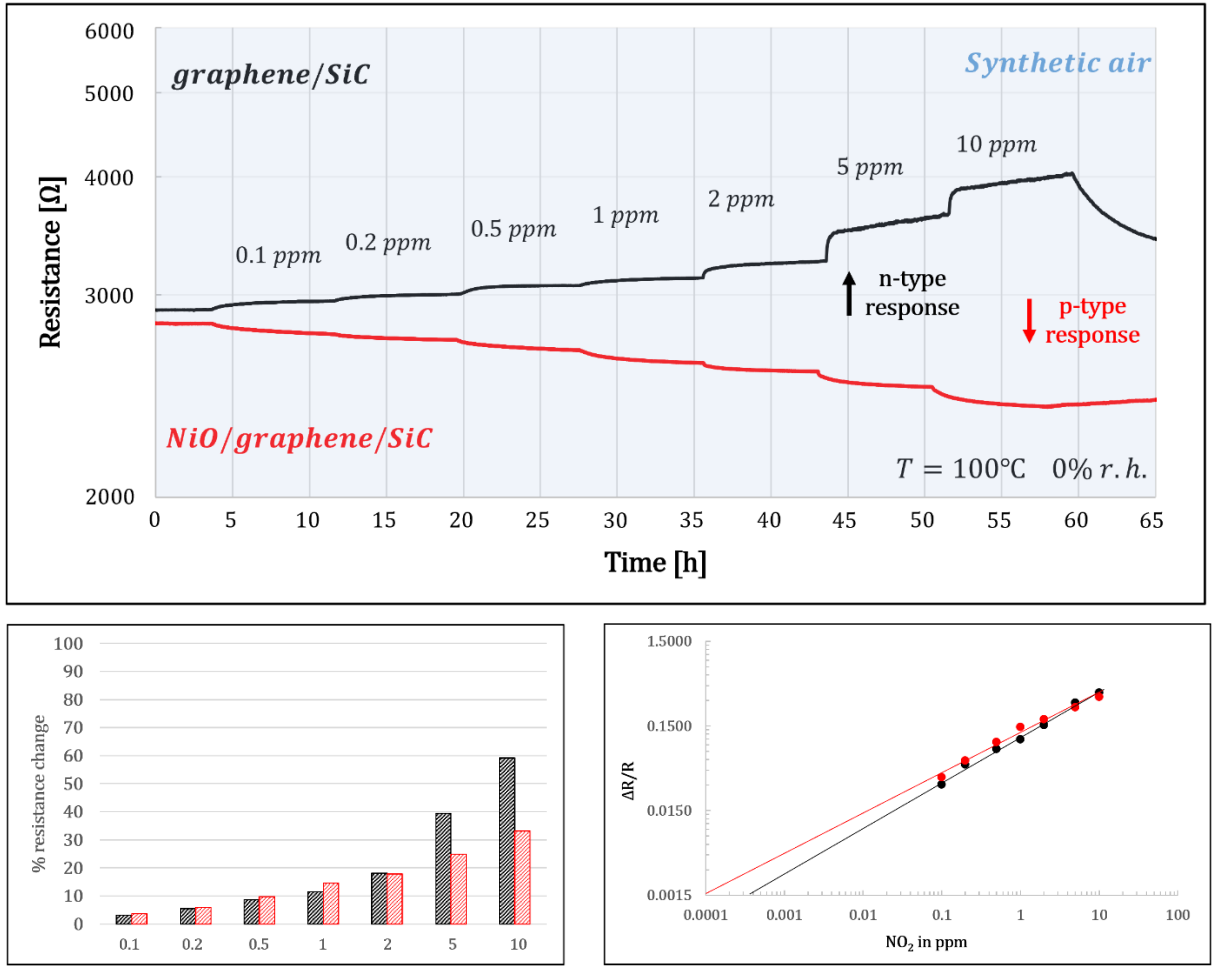

**Figure 10**: a) comparison between graphene/SiC and NiO/graphene/SiC sample towards $NO_2$, b) change in response towards $NO_2$ respect to the baseline, c) based on a power law response, the estimation of the potential detection limit is below 1 ppb for both samples.

## Conclusions

Epitaxial graphene/SiC and NiO/graphene/SiC samples were tested as gas sensors, specifically focusing on $NO_2$, where both showed very good responses, potentially opening the door for sub-ppb gas sensing. Upon exposure to $NO_2$, the NiO/graphene/SiC sample showed a non-reversible electrical conduction type switch from *n*-type to *p*-type. This behavior can be a consequence of the NiO layer, which lowers the Fermi level below the Dirac point of the graphene, changing its conduction type. If we consider that the NiO layer on top does not considerably alter sensitivity in comparison with the *n*-type graphene/SiC sample, this new NiO/graphene/SiC configuration can be an attractive sub-ppb *p*-type sensor.

## Acknowledgments

The work of F.S. was supported by a Georg Forster Fellowship from the Alexander von Humboldt Foundation. The work of S.S.N was supported by Ministry of Science, Research and Technology of Iran and University of Mazandaran. E.L.K. and G.L. acknowledge the support from the Helmholtz Energy Materials Foundry (HEMF) and PEROSEED (ZT-0024) project. This work was carried out in the framework of the Joint Lab GEN_FAB. The authors would also like to thank Prof. N. Koch for providing access the XPS equipment. Part of this work was performed in the framework of GraFOx, a Leibniz-ScienceCampus partially funded by the Leibniz association. M.B. gratefully acknowledges financial support by the Leibniz association.